\documentclass[journal, hidelinks]{IEEEtran}
\usepackage{amsfonts}
\usepackage{algorithmic}
\usepackage{algorithm}
\usepackage{array}
\usepackage{textcomp}
\usepackage{subfigure}
\usepackage{stfloats}
\usepackage{url}
\usepackage{verbatim}
\usepackage{graphicx}
\usepackage{lipsum}
\usepackage{cite}
\usepackage{orcidlink}
\usepackage{amsmath,amssymb,mathtools}
\usepackage[english]{babel}
\usepackage{caption}
\usepackage{bm}
\usepackage{csquotes}
\usepackage[acronym,shortcuts]{glossaries}
\usepackage{xcolor, flushend}
\usepackage{hyperref}

\newacronym{CDF}{CDF}{cumulative distribution function}
\newacronym{MIMO}{MIMO}{massive input massive output}
\newacronym{BS}{BS}{base station}
\newacronym{UE}{UE}{user equipment}
\newacronym{6G}{6G}{sixth-generation}
\newacronym{5G}{5G}{fifth-generation}
\newacronym{RIS}{RIS}{reconfigurable intelligent surface}
\newacronym{UPA}{UPA}{uniform planar array}
\newacronym{LS}{LS}{least squares}
\newacronym{ISAC}{ISAC}{integrated sensing and communications}
\newacronym{AoA}{AoA}{angle-of-arrival}
\newacronym{SINR}{SINR}{signal-to-interference-plus-noise ratio}
\newacronym{MRC}{MRC}{maximum ratio combining}
\newacronym{MUSIC}{MUSIC}{multiple signal classification}
\newacronym{CSI}{CSI}{channel state information}
\newacronym{LOS}{LOS}{line-of-sight}
\newacronym{NLOS}{NLOS}{non-line-of-sight}
\newacronym{AWGN}{AWGN}{additive white Gaussian noise}
\newacronym{CGA}{CGA}{conjugate gradient ascent}
\newacronym{NMSE}{NMSE}{Normalized Mean Square Error}

\begin{document}

\title{Blinding the Wiretapper: \\
RIS-Enabled User {Occultation} in the ISAC Era}

\author{Getuar Rexhepi,~\IEEEmembership{Student~Member,~IEEE}, Hyeon Seok Rou,~\IEEEmembership{Member,~IEEE}, \\Giuseppe Thadeu Freitas de Abreu,~\IEEEmembership{Senior~Member,~IEEE}, and\\ George C. Alexandropoulos,~\IEEEmembership{Senior~Member,~IEEE}
\thanks{G.~Rexhepi, H.~S.~Rou, and G.~T.~F. de~Abreu are with the School of Computer Science and Engineering, Constructor University Bremen, Campus Ring 1, 28759 Bremen, Germany (emails: [grexhepi, hrou, gabreu]@constructor.university).} 
\thanks{G.~C.~Alexandropoulos is with the Department of Informatics and Telecommunications, National and Kapodistrian University of Athens, 16122 Athens, Greece, and also with the Department of Electrical and Computer Engineering, University of Illinois Chicago, IL 60601, USA (e-mail: alexandg@di.uoa.gr).}
\vspace{-3.5ex}
}

\maketitle

\begin{abstract}
An undesirable consequence of the foreseeable proliferation of sophisticated \ac{ISAC} technologies is the enabling of spoofing, by malicious agents, of  situational information (such as proximity, direction or location) of legitimate users of wireless systems.
In order to mitigate this threat, we present a novel \ac{ISAC} scheme that, aided by a \ac{RIS}, enables the occultation of the positions of \ac{UE} from wiretappers, while maintaining both sensing and desired communication performance between the \acp{UE} and a legitimate \ac{BS}.
To that end, we first formulate an \ac{RIS} phase-shift optimization problem that jointly maximizes the sum-rate performance of the \acp{UE} (communication objective), while minimizing the projection of the wiretapper's effective channel onto the legitimate channel (hiding objective), thereby disrupting the attempts by a wiretapper of localizing the \acp{UE}.
Then, in order to efficiently solve the resulting non-convex joint optimization problem, a novel manifold optimization algorithm is derived, whose effectiveness is validated by numerical results, which demonstrate that the proposed approach preserves legitimate \ac{ISAC} performance while significantly degrading the wiretapper's sensing capability.
\end{abstract}

\begin{IEEEkeywords}
\Ac{RIS}, \ac{ISAC}, manifold optimization, sum-rate maximization, privacy, MUSIC.
\end{IEEEkeywords}

\glsresetall

\vspace{-2ex}
\section{Introduction}

In recent years, \acp{RIS} have gained significant attention for their remarkable ability to dynamically manipulate the incoming electromagnetic waves through their numerous reconfigurable elements, effectively reshaping the wireless environments in order to improve the performance of wireless systems via enhancing signal strength, extending coverage, mitigating interference, and more.\cite{Liu_CST21,Emil_RIS22,Basar2024}

Beyond improving communication performance, the development of \ac{RIS} technology also offers opportunities to strengthen other key performance criteria of wireless systems, such as the quality of sensing especially in the scope of \ac{ISAC} \cite{Shao2023, Elbir2023, Liu2023}, which is a crucial innovation of \ac{6G} technologies over \ac{5G} systems \cite{Alouini2024}.

However, given the inherently open nature of the wireless channels, the development and eventual proliferation of increasingly sophisticated \ac{ISAC} technologies capable of extracting situational and sensitive information about users even from encrypted signals \cite{RayanWCNC2025}, is poised to significantly widen the attack space of \ac{6G} wireless systems \cite{Mucchi2021}.

This fact has prompted the research community to investigate secure \ac{ISAC} techniques\cite{He2024, Zou2024,Boroujeni2025}, and although much of the \ac{RIS} literature concentrated mostly on enhancing communications performance at first, and latter considering also sensing accuracy, but often overlooking the vital issue of protection against wiretapping and eavesdropping, works aiming to explore \ac{RIS}-based solutions to offer physical layer security have also started to appear \cite{Kaur2024, Li2025, George2023, Yang2024, magbool2025hiding}.
Excellent examples are \cite{George2023}, where \ac{RIS}-aided safeguarding solutions against eavesdropping boosted with a malicious \ac{RIS} under different levels of channel information availability were offered; the scheme in \cite{Yang2024}, which seeks to design sensing signals such that they also function as artificial noise to potential eavesdroppers; and the work in \cite{magbool2025hiding}, where a downlink \ac{RIS}-assisted user-location shielding method was presented. 

\begin{figure}[H]
\centering
\includegraphics[width=\columnwidth]{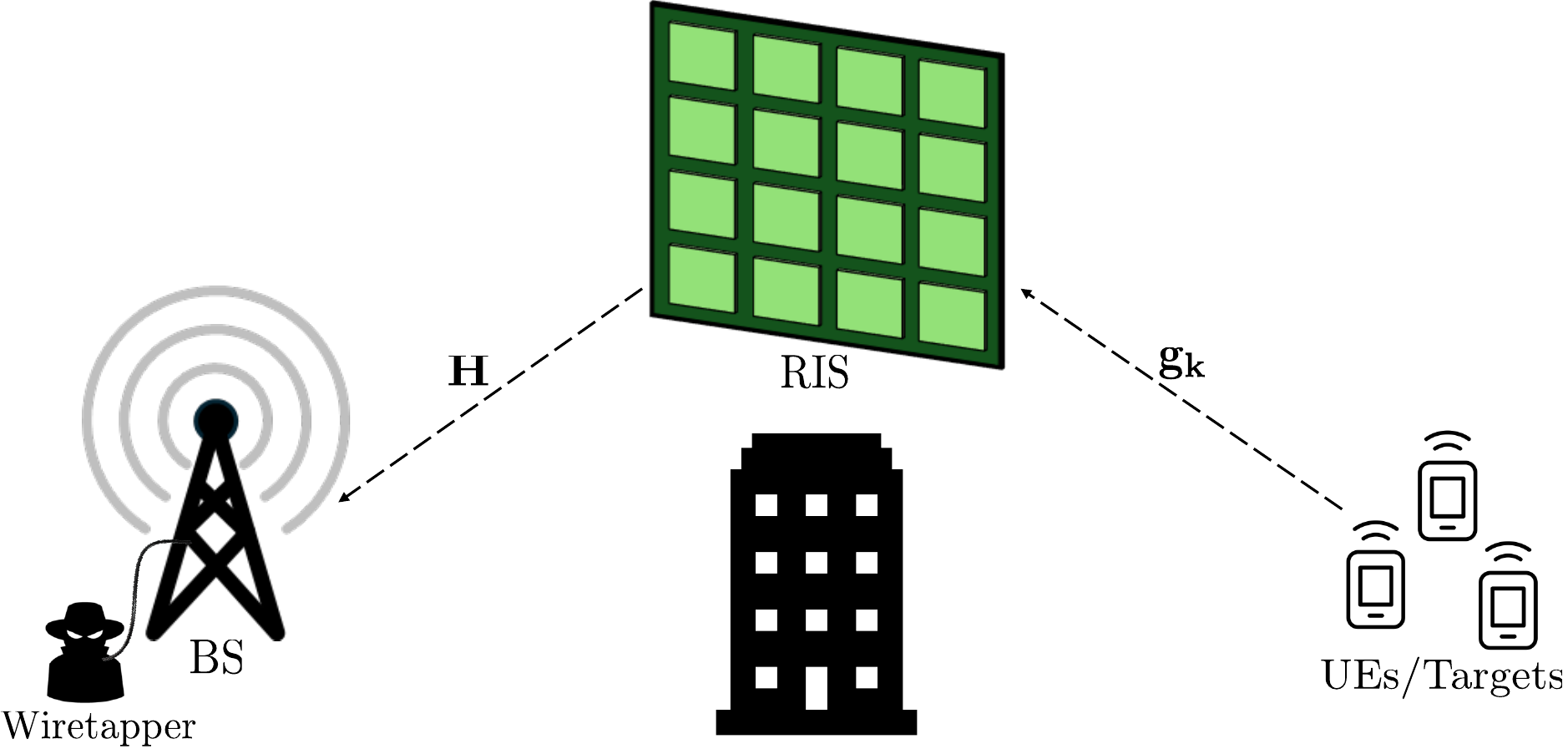}
\caption{The \ac{RIS}-assisted Uplink secure \ac{ISAC} system.}    
\label{fig:system_model}
\vspace{-2ex}
\end{figure}

In this paper, we contribute to this emerging topic of \ac{RIS}-aided secure \ac{ISAC} system design, with a novel uplink \ac{ISAC} scheme in which an \ac{RIS} is optimized to maximize the sum rate achievable by a set of \acp{UE}, while preventing a wiretapper physically connected to the \ac{BS} antennas from extracting information about their location.
The considered scenario, illustrated in Figure \ref{fig:system_model}, is therefore such that the wiretapper is assumed to have full access to the entire receive signal without any additional distortion.
Under such conditions, the only leverage that the \ac{BS} has over the wiretapper with respect to sensing, $i.e.$, the estimation of the \acp{AoA} and distances of the \acp{UE}, is its ability to parameterize the \ac{RIS} so as to alter the spatial signatures of the \acp{UE}, preventing the wiretapper's sensing attack.

To that end, the proposed method is based on the formulation of a manifold optimization problem through which the phase shifts of \ac{RIS} elements are adjusted to simultaneously ensure high communication rate and the desired occultation of the sensing information.
Simulations results validate the effectiveness of the proposed approach, showcasing accurate sensing by \ac{BS} and a significantly deteriorated sensing performance by the wiretapper, even if the latter is assumed to have full knowledge of the structure of the \ac{RIS}.
In addition, a significantly improved sum-rate performance over the non-optimized system is demonstrated, consolidating the effectiveness of the proposed \ac{ISAC} method.

\vspace{-1ex}
\section{System and Channel Models}
\label{sec:system_model}

Consider an \ac{RIS}-assisted uplink \ac{ISAC} scenario between a multi-antenna \ac{BS} equipped with $M$ receive antennas, and a set of $K$ single-antenna \acp{UE}.
The \ac{RIS} deployed in the environment is modeled as a \ac{UPA} with $N$ elements, which is controlled by the \ac{BS}\footnotemark.

\footnotetext{This implies the assumption that the control signals transmitted from the \ac{BS} to the \ac{RIS} are encrypted and cannot be decoded by the wiretapper.}

It is assumed that the direct \ac{LOS} path between the \ac{BS} and \acp{UE} is blocked by obstacles, such that the \ac{NLOS} channel via the \ac{RIS} is the available medium for communications and localization.
The planar \ac{RIS} in the YZ-plane is composed of $N_\mathrm{h}$ horizontal and $N_\mathrm{v}$ vertical elements, with $N \triangleq N_\mathrm{h} N_\mathrm{v}$ total elements, with inter-element spacings $d_\mathrm{h}$ and $d_\mathrm{v}$, respectively.
Assuming that both $N_\mathrm{h}$ and $N_\mathrm{v}$ are odd\footnote{This is without loss of generality, and adopted only so that there is a geometrically-central row and column which simplifies the notation. The proposed method can, however, be extended to even numbers with correspondingly modified indices.}, the horizontal and vertical indices of the \ac{RIS} elements are denoted by $N_\mathrm{h}$ and $N_\mathrm{v}$, where $n_\mathrm{h} = -\frac{N_\mathrm{h}-1}{2}, \dots, 0, \dots, \frac{N_\mathrm{h}-1}{2},$ and $n_\mathrm{v} = -\frac{N_\mathrm{v}-1}{2}, \dots, 0, \dots, \frac{N_\mathrm{v}-1}{2}$, such that the element at the geometrical center of the \ac{RIS} is located at $(n_\mathrm{h}, n_\mathrm{v}) = (0, 0)$. Given that the global axis is defined such that the center of the \ac{RIS} is located at the origin, the 3D location of the $k$-th \ac{UE} ($k=1,\ldots,K$) is described by
\begin{equation}
\label{eq:location}
\begin{pmatrix}
x_k \\
y_k \\
z_k \\
\end{pmatrix}
\triangleq
\begin{pmatrix}
r_k \cos(\varphi_k)\cos(\theta_k) \\
r_k \sin(\varphi_k)\cos(\theta_k) \\
r_k \sin(\theta_k) \\
\end{pmatrix},
\end{equation}
where $r_{k} \in \mathbb{R}$, $\varphi_{k} \in [-\pi, +\pi)$, and $\theta_{k} \in [-\pi, +\pi)$ represent the distance and the azimuth and elevation \ac{AoA} between the $k$-th \ac{UE} and the center of the \ac{RIS}, respectively.

From the above, it follows that the distance between the $(n_\mathrm{h}, n_\mathrm{v})$-th \ac{RIS} element and the $k$-th \ac{UE} is given by
\begin{equation}
\label{eq:distance}
~\!r_{k}^{\left(n_\mathrm{h}, n_\mathrm{v}\right)} \!=\! \sqrt{\left(x_k \right)^{2} \!+\! \left(y_k - n_\mathrm{h} d_\mathrm{h}\right)^{2} \!+\! \left(z_k - n_\mathrm{v} d_\mathrm{v}\right)^{2}~},
\end{equation}
which, under the Fresnel approximation, can be expressed as
\begin{align}
 r_{k}^{\left(n_\mathrm{h}, n_\mathrm{v}\right)} \approx &\; r_{k} - n_\mathrm{h} d_\mathrm{h} \sin (\varphi_{k}) \cos (\theta_{k}) - n_\mathrm{v} d_\mathrm{v} \sin (\theta_{k})\nonumber\\
\label{eq:distance_approx}
& + \frac{n_\mathrm{h}^2 d_\mathrm{h}^2 + n_\mathrm{v}^2 d_\mathrm{v}^2}{2r_{k}}.
\end{align}

Finally, for future convenience, we collect the $N$ tunable \ac{RIS} phase shifts $\phi_n$, with $n \in \{1,\dots,N\}$, into the diagonal matrix
\begin{equation}
\!\mathbf{\Phi} \triangleq \text{diag}(\bm{\phi}), \quad\! \text{with}\quad\!
\bm{\phi} \triangleq [\phi_1, \ldots, \phi_N]^\mathsf{T} \in \mathbb{C}^{N \times 1},\!\!
\label{eq:RIS_phase_shift_matrix}
\end{equation} 
where $\phi_n\triangleq e^{j\pi\varphi_n}$, with $\varphi_n \in [0,2\pi)$, and $n \triangleq (n_\mathrm{v} + \frac{N_\mathrm{v}-1}{2})\cdot N_\mathrm{h} + (n_\mathrm{h} + \frac{N_\mathrm{h}-1}{2})$, $i.e.$, counted left-to-right and bottom-to-top.

\subsection{Signal and Channel Models}

During the synchronized uplink communication at time slot $t \in \{1,\ldots, T\}$, each $k$-th \ac{UE} transmits the sequence $s_k(t) \in \mathbb{C}$, assumed to be zero mean with $\mathbb{E}\{s_k^{\vphantom{*}}(t) s_k^*(t)\} = 1$. Given that each \ac{UE} transmits with power $p_k$, the received signal $\mathbf{y}(t) \in \mathbb{C}^{M \times 1}$ at the \ac{BS} can be expressed as
\vspace{-0.5ex}
\begin{equation}
\mathbf{y}(t) = \sum_{k=1}^{K} \sqrt{p_k}\, \mathbf{H}\mathbf{\Phi}\mathbf{g}_k s_k(t) + \mathbf{z}(t), 
\vspace{-0.5ex}
\end{equation}
where $\mathbf{z}(t) \sim \mathcal{CN}(\mathbf{0}, \sigma^2_\mathrm{z}\mathbf{I}_M) \in \mathbb{C}^{M \times 1}$ is the \ac{AWGN} vector, $\mathbf{H} \in \mathbb{C}^{M \times N}$ is the far-field channel matrix from the \ac{RIS} to the \ac{BS}, modeled as a Rician fading channel with a factor $K = 2$, and $\mathbf{g}_k \in \mathbb{C}^{N \times 1}$ is the near-field channel vector between the \ac{RIS} and each $k$-th \ac{UE}, which is modeled as
\vspace{-0.5ex}
\begin{equation}
\label{eq:near_field_channel_model}
\mathbf{g}_k = \frac{\lambda}{4\pi r_{k}}\mathbf{a}_R\big(r_k, \varphi_k, \theta_k\big),
\vspace{-0.5ex}
\end{equation}
%
%
%
where $\mathbf{a}_R(r_k, \varphi_k, \theta_k)$ is the \ac{RIS} array response vector towards each $k$-th \ac{UE} which, at the $(n_\mathrm{h}, n_\mathrm{v}$)-th element, is given by
\vspace{-1.5ex}
\begin{equation}
\begin{aligned}\label{eq:array_response}
\big[\mathbf{a}_R(r_k, \varphi_k, \theta_k)\big]_{(n_\mathrm{h}, n_\mathrm{v})} = e^{\,j \frac{2 \pi}{\lambda}\!\left(r_{k}-r_{k}^{\left(n_\mathrm{h}, n_\mathrm{v}\right)}\right)},
\end{aligned}
\end{equation}
where $r_{k}^{\left(n_\mathrm{h}, n_\mathrm{v}\right)}$ is the \ac{RIS} element-dependent distance given by~\eqref{eq:distance} and $\lambda$ is the wavelength.


The received signal can be formatted as an effective matrix  representing all users and discrete time slots, by first collecting the transmit signals $\tilde{s}_k = \sqrt{p_k \eta_k} $ of the $K$ \acp{UE} at all $T$ discrete time indices\footnote{It is assumed that the channel statistics remain static during the total transmission period $T$.}, yielding
\begin{equation}
\mathbf{S} \triangleq
\begin{bmatrix}
\tilde{s}_1(1) & \cdots & \tilde{s}_1(T) \\
\vdots & \ddots & \vdots \\
\tilde{s}_K(1) & \cdots & \tilde{s}_K(T) \\
\end{bmatrix} \in \mathbb{C}^{K \times T}.
\end{equation} 

By leveraging the Fresnel approximation of the \ac{RIS}-element distances in~\eqref{eq:distance_approx}, the effective received signal can also be compactly formulated as
\begin{equation}
\label{eq:received_signal_matrix}
\mathbf{Y} \triangleq [\mathbf{y}(1),\ldots,\mathbf{y}(T)] = \mathbf{H} \mathbf{\Phi} \mathbf{A} \mathbf{S} + \mathbf{Z} \in \mathbb{C}^{M \times T},
\end{equation}
where $\mathbf{Z} \triangleq [\mathbf{z}(1),\ldots,\mathbf{z}(T)] \in \mathbb{C}^{M \times T}$ is the effective \ac{AWGN} matrix, while $\mathbf{A} \in \mathbb{C}^{N \times K}$ is the effective array response matrix  between the \acp{UE} and the \ac{RIS}, such that its $(n,k)$-th element is given by
\begin{align}
a_{n,k} & = \big[\mathbf{a}_R(r_k, \varphi_k, \theta_k)\big]_{(n_\mathrm{h}, n_\mathrm{v})}\\[1ex]
& \approx \exp\Big(\!-j\big(n_\mathrm{h} \alpha_k + n_\mathrm{v} \beta_k  + \big(n_\mathrm{h}^2 d_\mathrm{h}^2 + n_\mathrm{v}^2 d_\mathrm{v}^2\big)\gamma_k \big)\Big)
\nonumber
\end{align}
with the auxiliary coefficients $\alpha_k \triangleq \frac{2\pi}{\lambda} d_\mathrm{h} \sin(\varphi_k)\cos(\theta_k)$, $\beta_k \triangleq \frac{2\pi}{\lambda} d_\mathrm{v} \sin(\theta_k)$, and $\gamma_k \triangleq \frac{\pi}{\lambda r_{k}}$ derived from the array response in~\eqref{eq:array_response} and the Fresnel approximation in~\eqref{eq:distance_approx}.

\vspace{-1ex}
\section{Secure ISAC Problem Formulation}

Next, formulate the optimization problem for the parameterization of \ac{RIS} phase matrix $\mathbf{\Phi}$ aimed at achieving the dual objective of enhancing the communications performance, while enabling the occultation of the \acp{UE} from a potential wiretapper. 

%
%

\vspace{-1ex}
\subsection{Occultation Objective}

\setcounter{footnote}{\thefootnote - 3}

We make the worst-case assumption that the wiretapper has access to the exact received signal $\mathbf{Y}$, as well as perfect knowledge of both the \ac{RIS}-to-\ac{BS} channel matrix $\mathbf{H}$ and the structure of the array response matrix $\mathbf{A}$, but \underline{not} the phase configuration matrix $\mathbf{\Phi}$\footnotemark.
In other words, the wiretapper is able to reconstruct the effective channel
\vspace{-0.5ex}
\begin{equation}
\label{eq:WT_effective_channel}
\mathbf{G}_W \triangleq \mathbf{H}\mathbf{\Phi}_W\mathbf{A} \in \mathbb{C}^{M \times K},
\vspace{-0.5ex}
\end{equation}
where $\mathbf{\Phi}_W$ is the wiretappers own guess of the \ac{RIS} phase configuration matrix.

In this work, we will hereafter assume that in absence of any knowledge the configuration of the \ac{RIS}, the wiretapper sets $\mathbf{\Phi}_W=\mathbf{I}_{N\times N}$, highlighting however that the particular choice has no fundamental implication onto the performance of the proposed method, and leaving further investigation for
a follow-up work.

In contrast to the wiretapper, the \ac{BS} is able to design $\mathbf{\Phi}$, such that its effective channel is described by
\vspace{-0.5ex}
\begin{equation}
\label{eq:BS_effective_channel}
\mathbf{G}_B \triangleq \mathbf{H}\mathbf{\Phi}\mathbf{A} \in \mathbb{C}^{M \times K},
\vspace{-0.5ex}
\end{equation}
such that the system's strategy to hide the spatial signatures of the \acp{UE} from the wiretapper is to  minimize the projection of the $\mathbf{G}_W$ onto $\mathbf{G}_B$, that is, to ensure that $\left\| \mathbf{G}^\mathsf{H}_B \mathbf{G}_W \right\|^2_\mathrm{F} \leq \epsilon$, for a sufficiently small $\epsilon$.


\setcounter{footnote}{\thefootnote + 2}

\vspace{-1ex}
\subsection{Communications Objective}

In light of the above, the effective channel between the \ac{BS} and the $k$-th \ac{UE} is given by $\mathbf{v}_k \triangleq \mathbf{H}\mathbf{\Phi}\mathbf{g}_k \in \mathbb{C}^{M \times 1}$, such that after maximum-ratio combining via $\mathbf{w}_k = \frac{\mathbf{v}_k}{\|\mathbf{v}_k\|}$, the estimated signal corresponding to the $k$-th \ac{UE} at discrete time index $t$ is given by $\hat{s}_k(t) = \mathbf{w}_k^\mathsf{H}  \mathbf{y}(t) \in \mathbb{C}$ with \ac{SINR} 
\vspace{-0.5ex}
\begin{equation}
\mathrm{SINR}_k = \frac{p_k |\mathbf{w}_k^\mathsf{H} \mathbf{v}_k|^2}{\sum\limits_{j \neq k} p_j |\mathbf{w}_k^\mathsf{H} \mathbf{v}_j|^2 + \sigma_\mathrm{z}^2 \|\mathbf{w}_k\|^2}.
\vspace{-0.5ex}
\end{equation}

The communication objective of the \ac{BS} is to design $\mathbf{\Phi}$ so as to maximize the total achievable rate over all \acp{UE}, $i.e.$
\vspace{-0.5ex}
\begin{equation}
\mathrm{R} \triangleq \sum_{k=1}^K \mathrm{R}_k,\quad \text{with}\quad \mathrm{R}_k = \log_2 \left( 1 + \text{SINR}_k \right).
\vspace{-0.5ex}
\end{equation}

\subsection{Joint Optimization Problem}

Under all the above, the optimization problem to be solved by the \ac{BS} can be described by 
\begin{equation}
\begin{aligned}
& \underset{\mathbf{\Phi}}{\text{max}}
& & \sum_{k=1}^{K} \log_2\left(1+\text{SINR}_k\right)\\[1ex]
& \text{subject to}
& & |\phi_n| = 1\, \forall n,\,\,\left\| \mathbf{G}^\mathsf{H}_B \mathbf{G}_W \right\|^2_\mathrm{F} \leq \epsilon,
\end{aligned}
\end{equation}
which can be reformulated by moving the projection constraint of the effective channels into the objective function, yielding
\begin{equation}
\label{eq:joint_objective}
\begin{aligned}
& \underset{\mathbf{\Phi}}{\text{max}}
& & \rho\sum_{k=1}^{K} \log_2\left(1+\text{SINR}_k\right) + (1-\rho) \Gamma\\[1ex]
& \text{subject to}
& & |\phi_n| = 1\, \forall n,
\end{aligned}
\end{equation} 
where $\Gamma \triangleq \max\{{\left\| \mathbf{G}^\mathsf{H}_B \mathbf{G}_W \right\|^2_\mathrm{F} - \epsilon, 0}\}$ and $\rho \in [0, 1]$ is a trade-off parameter of choice.
\newpage

The optimization problem described in equation \eqref{eq:joint_objective} is non-convex due to the \ac{RIS} constraint to the unit modulus, namely $|\phi_n| = 1~\forall n$, but can be solved efficiently using manifold optimization techniques \cite{boumal2023}.
To that end, we derive in the sequel the corresponding \ac{CGA} algorithm with the gradient of the objective function in equation \eqref{eq:joint_objective} offered in closed form.

\subsection{Proposed Solution via CGA over the Riemannian Manifold}

Straightforwardly, the gradient of the objective function in problem \eqref{eq:joint_objective} with respect to the configuration vector $\bm{\phi}$ is
\begin{equation}\label{eq:gradient}
\begin{aligned}
\nabla_{\!\bm{\phi}} \mathcal{L}(\bm{\phi}) &= \nabla_{\!\bm{\phi}} \bigg( \sum_{k=1}^{K} \log_2\left(1+\text{SINR}_k\right) + \rho \Gamma \bigg)\\
&= \sum_{k=1}^{K} \frac{1}{\ln(2)} \frac{\nabla_{\!\bm{\phi}} \text{SINR}_k}{1+\text{SINR}_k} + \rho \nabla_{\!\bm{\phi}} \Gamma,
\end{aligned}
\end{equation}
where
\begin{eqnarray}
\label{eq:SINR_grad}
&\nabla_{\!\bm{\phi}} \text{SINR}_k = \frac{\delta_k  \nabla_{\!\bm{\phi}} \nu_k - \nu_k \nabla_{\!\bm{\phi}} \delta_k}{\delta_k^2}&\\
&\nabla_{\!\bm{\phi}} \Gamma = 
\begin{cases}
\displaystyle \mathbf{H}^\mathsf{H}\mathbf{G}_W \mathbf{G}_W^\mathsf{H} \mathbf{G}_B \mathbf{A}^\mathsf{H}, & \left\| \mathbf{G}^\mathsf{H}_B \mathbf{G}_W \right\|^2_\mathrm{F}> \epsilon \\[2mm]
\mathbf{0}, \quad \text{otherwise},
\end{cases}&\\
\label{eq:gradient_nu}
&\nabla_{\!\bm{\phi}} \nu_k = 2 {p}_k \left(\text{diag}(\mathbf{g}_k^\mathsf{H}) \mathbf{H}^\mathsf{H} \mathbf{H} \text{diag}(\mathbf{g}_k) \right) \bm{\phi} ,& \\
\label{eq:gradient_delta}
&\nabla_{\!\bm{\phi}} \delta_k = \sum\limits_{j \neq k}2 p_j \left(\text{diag}(\mathbf{g}_j^\mathsf{H}) \mathbf{H}^\mathsf{H} \mathbf{w}_k \mathbf{w}_k^\mathsf{H} \mathbf{H} \text{diag}(\mathbf{g}_j) \right) \bm{\phi} ,& \\
\label{eq:nu}
&\nu_k = p_k |\mathbf{w}_k^\mathsf{H} \mathbf{v}_k|^2,&\\
\label{eq:delta}
&\delta_k = \sum\limits_{j \neq k} p_j |\mathbf{w}_k^\mathsf{H} \mathbf{v}_j|^2 + \sigma_\mathrm{z}^2 \|\mathbf{w}_k\|^2.&
\end{eqnarray}

In the context of first-order optimization methods, the steepest ascent algorithm represents the most fundamental approach for locating a local maximum of a non-convex function.
However, instead of relying solely on the steepest ascent, we exploit the underlying geometry of the problem to approximate second-order information by comparing first-order data (tangent vectors) at different points on the manifold.

For the problem at hand, the manifold of interest is a Riemannian manifold consisting of the complex unit circle in $\mathbb{C}^N$, defined as the set of complex vectors with unit modulus, $i.e.$, $\mathcal{M} = \left\{ \mathbf{x} \in \mathbb{C}^{N \times 1} : |\mathbf{x}| = 1 \right\}$.
Consequently, steepest ascent is performed via a succession of a retraction operation ${R}(\cdot,\cdot)$, followed by a projection ${T}(\cdot,\cdot)$ onto the tangent space, which for the complex unit circle Riemannian manifold are respectively given by
\begin{equation}
{R}(\mathbf{x}, \boldsymbol{\xi}) = \frac{\mathbf{x} + \alpha \boldsymbol{\xi}}{|\mathbf{x} + \alpha \boldsymbol{\xi}|}, \label{eq:retraction}
\end{equation}
and
\begin{equation}
T_{\mathbf{x}}( \nabla_{\mathbf{x}}F, \mathbf{x} ) = \nabla_{\mathbf{x}}F - \Re\{ \nabla_{\mathbf{x}}F \otimes \mathbf{x}^\mathsf{H} \} \otimes \mathbf{x}, \label{eq:projection}
\end{equation}
where $\alpha$ denotes the step size, $\mathbf{x}$ is the current point on the manifold, $\boldsymbol{\xi}$ is the ascent direction in the tangent space, and $\nabla_{\mathbf{x}}F$ represents the gradient of the objective function at $\mathbf{x}$.

The complete \ac{CGA} algorithm is outlined in Algorithm~\ref{alg:conjugate_gradient}, which begins with a randomly initialized set of phase shifts $\boldsymbol{\phi}^{(0)}$, only constrained to lie on the complex unit circle.

\begin{algorithm}[H]
\caption{Proposed Privacy-preserving RIS Optimization}
\label{alg:conjugate_gradient}
\begin{algorithmic}[1]
\STATE \textbf{Input:} $\mathbf{H}$, $\mathbf{A}$, $K$, $M$, $N$, $\rho$, and $\epsilon$.
\STATE \textbf{Initialize:} $\mathbf{\Phi}^{(0)} = \text{diag}(\bm{\phi}^{(0)}) \in \mathcal{M}$.  \\
\STATE \textbf{Compute: }$\mathcal{L}(\bm{\phi}^{(0)})$, $\mathbf{r}^{(0)} = \nabla_{\!\bm{\phi}}\mathcal{L}(\bm{\phi}^{(0)})$, and $\bm{\xi}^{(0)} = \mathbf{r}^{(0)}$.     
\STATE \textbf{while} not converged \textbf{do}
\STATE \quad \textbf{if} $\langle \mathbf{r}^{(i)}, \bm{\xi}^{(i)} \rangle \leq 0 $ \textbf{then}
\STATE \quad \quad Set $\bm{\xi}^{(i)} = \mathbf{r}^{(i)}$.
\STATE \quad \textbf{end if}
\STATE  \quad Compute $\alpha^{(i)}$ using Armijo line search.
\STATE  \quad Set $\bm{\phi}^{(i+1)} = R(\bm{\phi}^{(i)}, \alpha^{(i)} \bm{\xi}^{(i)})$ using (\ref{eq:retraction}).
\STATE  \quad Set $\mathbf{r}^{(i+1)}=\mathcal{L}(\bm{\phi}^{(i+1)})$ using (\ref{eq:gradient}).
\STATE  \quad Compute $\mathbf{r}^{(i)}_{\dagger} = T_{\!\bm{\phi}} (\mathbf{r}^{(i)}, \bm{\phi}^{(i)})$ using (\ref{eq:projection}).
\STATE  \quad Set $\beta^{(i)} = \max(0, \langle \mathbf{r}^{(i)} - \mathbf{r}^{(i)}_{\dagger}, \mathbf{r}^{(i+1)} \rangle / \langle \mathbf{r}^{(i)}, \bm{\xi}^{(i)} \rangle)$.
\STATE  \quad Calculate $\bm{\xi}^{(i+1)} = \mathbf{r}^{(i+1)} + \beta^{(i)} \bm{\xi}^{(i)}$.
\STATE \textbf{end while}
\STATE \textbf{Output:} $\mathbf{\Phi} = \mathrm{diag}(\bm{\phi}^{(\mathrm{conv})})$.
\end{algorithmic}
\end{algorithm}
\vspace{-1ex}

At each iteration, the algorithm evaluates the objective function and its gradient, and updates the phase shifts accordingly until convergence is achieved.
To determine the step size $\alpha$, an Armijo line search can be employed~\cite{nocedal2006}, ensuring sufficient ascent in each iteration.
To mitigate oscillatory behavior, a restart mechanism is also incorporated, $i.e.$, if the inner product between consecutive gradients is negative, the search direction is reset to the current gradient.
Finally, a momentum term computed using the Polak-Ribi{\`e}re formula \cite{nocedal2006} is integrated into the gradient update to enhance convergence speed.

\vspace{-1ex}
\section{Simulation Results}

In this section, the effectiveness of the proposed \ac{RIS}-assisted secure \ac{ISAC} framework in achieving the dual objectives of enhancing communications performance and enabling \ac{UE} occultation is assessed empirically via computer simulations, executed with the following parameters.
The \ac{BS} is equipped with $M=128$ antennas, the \ac{RIS} consists of $N=121$ elements arranged in a $11 \times 11$ \ac{UPA}, and the system serves $K=3$ \acp{UE}. 
The inter-element spacing of the RIS is set to $d_\mathrm{h} = d_\mathrm{v} = 0.5\lambda$ with $\lambda = 0.3$ m. 
Each \ac{UE} transmittes with a power of $p_k = 10$ dBm, while the noise power is set to $\sigma^2 = -104$ dBm. 
The azimuth and elevation angles of the \acp{UE} are randomly generated within the range $[-\pi, \pi]$, and their distances are uniformly distributed within $[1, 20]$ meters.

\begin{figure}[H]
\centering
\subfigure[{\footnotesize AoA: Base Station}]%
{\includegraphics[width=0.975\columnwidth]{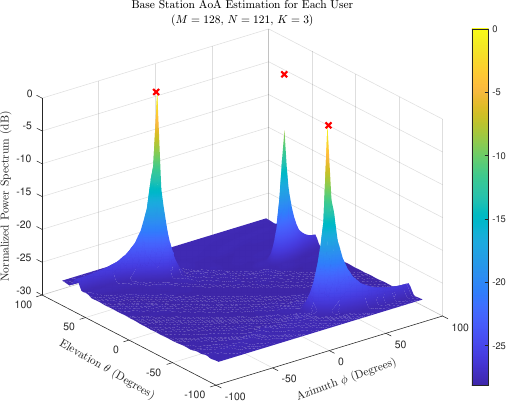}
\label{fig:MUSIC_BS}}\\
\subfigure[{\footnotesize AoA: Wiretapper}]%
{\includegraphics[width=0.975\columnwidth]{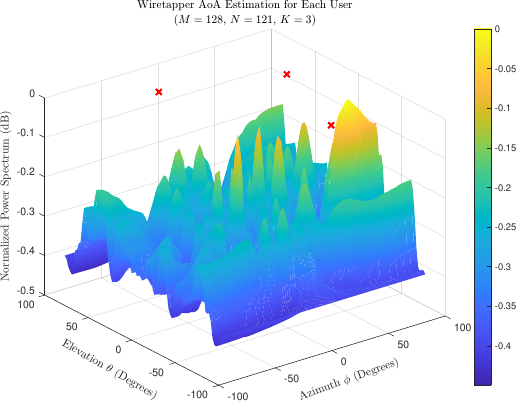}
\label{fig:MUSIC_WT}}
\subfigure[{\footnotesize Distances: BS and Wiretapper}]%
{\includegraphics[width=0.975\columnwidth]{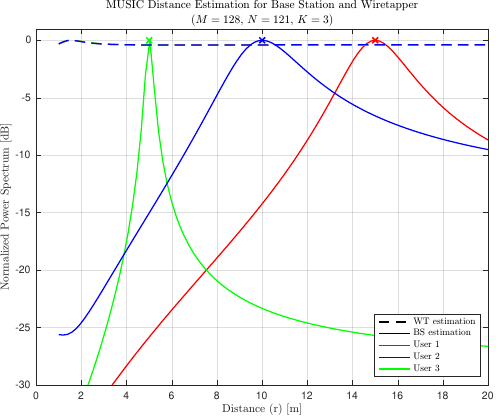}
\label{fig:distance}}
\vspace{-2ex}
\caption{MUSIC spectra obtained by the legitimate \ac{BS} and the wiretapper for the estimation of the \ac{UE} \acp{AoA} and distances.}
\label{fig:MUSICSpectra}
\end{figure}

Our first set of results are offered in Figure \ref{fig:MUSICSpectra}, which shows the \ac{MUSIC} spectra for the  estimation of the the \acp{AoA} and distances corresponding to the three \acp{UE}.
It can be observed that, with the proposed RIS design, the wiretapper becomes incapable to accurately detect the location of the \acp{UE}, due to the rough \ac{AoA} and flat distance spectra, respectively.
In contrast, the results show that the \ac{MUSIC} spectra obtained by the \ac{BS} is sufficient to successfully retrieve the \ac{AoA} and distances of all \acp{UE}.

\begin{figure}[H]
\centering
\includegraphics[width=\columnwidth]{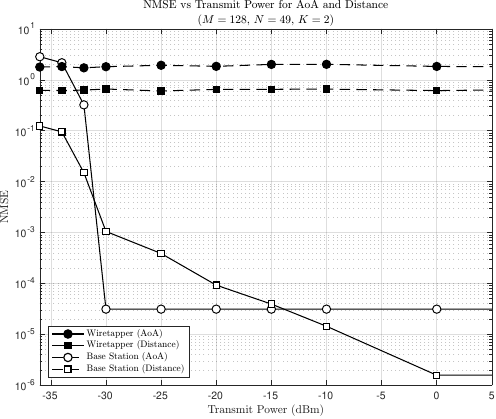}
\caption{NMSE performance.}
\label{fig:NMSE}
\includegraphics[width=\columnwidth]{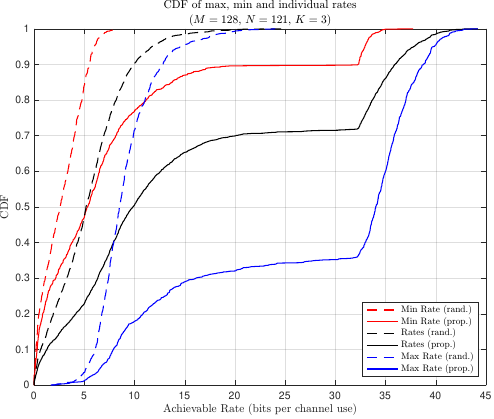}
\caption{Achievable sum-rate performance.}
\label{fig:Rate}
\vspace{-2ex}
\end{figure}

In order to complement the assessment of sensing performance, we further shown in Figure \ref{fig:NMSE} the \ac{NMSE} of the estimates obtained via the \ac{MUSIC} algorithm at the \ac{BS} and the wiretapper under varying transmit power levels.
It can be observed from the flat NMSE curves corresponding to the wiretapper that the scheme offers effective privacy to the \acp{UE} irrespective of the transmit power.

Finally, we evaluate the communication performance of the proposed method in terms of achievable rate.
In particular, we compare in Figure \ref{fig:Rate} the \acp{CDF} corresponding to the highest, lowest and average rates achieved by the \ac{BS} and the wiretapper.
It is found that the proposed \ac{RIS} optimization significantly enhances the sum rate compared to a non-optimized \ac{RIS} configuration, confirming the framework's capability to maintain high communications performance, while ensuring \ac{UE} occultation.

\vspace{1ex}
\section{Conclusion}

We presented a novel \ac{RIS}-assisted privadcy-preserving \ac{ISAC} scheme that leverages the reconfigurability of \acp{RIS} to conceal \acp{UE} from a wiretapper physically connected to the antennas of the \ac{BS} and therefore with full access to the received signal and \ac{RIS}-to-\ac{BS} channel.
To this end, we formulated a non-convex optimization problem to find the \ac{RIS} phase configuration that maximizes the \acp{UE}'s sum rate, while minimizing the projection of the wiretapper's effective channel onto the legitimate channel.
The problem was then solved via a Riemannian manifold steepest ascent technique with gradient computed in closed-form.
Simulation results validated the effectiveness of the proposed approach, demonstrating significant degradation in the wiretapper's sensing and communication capability, while preserving high communications and sensing performance at the legitimate \ac{BS}.

\end{document}